\documentclass[twocolumn,showpacs,prl]{revtex4}%
\usepackage{amssymb}
\usepackage{graphicx}
\usepackage{dcolumn}
\usepackage{bm}
\usepackage{amsmath}
\usepackage{amsfonts}%
\providecommand{\U}[1]{\protect\rule{.1in}{.1in}}

\begin{document}
\title{Supersolid and charge density-wave states from anisotropic interaction \\
in an optical lattice}
\author{Y.-H. Chan, Y.-J. Han, and L.-M. Duan}
\affiliation{Department of Physics and MCTP, University of Michigan, Ann Arbor, Michigan
48109 }
\affiliation{}
\date{\today }

\begin{abstract}
We show anisotropy of the dipole interaction between magnetic atoms or polar molecules
can stabilize new quantum phases in an optical lattice. Using a well
controlled numerical method based on the tensor network algorithm, we
calculate phase diagram of the resultant effective Hamiltonian in a
two-dimensional square lattice --- an anisotropic Hubbard model of hard-core
bosons with attractive interaction in one direction and repulsive interaction
in the other direction. Besides the conventional superfluid and the Mott insulator
states, we find the striped and the checkerboard charge density wave states
and the supersolid phase that interconnect the superfluid and the striped solid
states. The transition to the supersolid phase has a mechanism different from
the case of the soft-core Bose Hubbard model.

\end{abstract}
\maketitle

The remarkable experimental realization of ultracold dipolar molecules
\cite{1} and ultracold atoms with large magnetic moments \cite{2} open up the
possibilities of probing novel phases of matter that are induced by strong
dipole-dipole interaction. Among these possibilities, a particularly
interesting one is the supersolid phase, where superfluidity coexists with a
charge density wave order \cite{3}. In a square lattice, the supersolid state
has been predicted to exist for soft-core bosons if the off-site interaction
gets comparable with the on-site interaction \cite{4}. For ultracold atoms, it
is challenging to experimentally realize this condition, as even with the
long-range dipole interaction, the interaction strength still falls off
quickly with distance. In an optical lattice, the on-site interaction is
typically much larger than the off-site interaction, and if one increases the
interaction strength (or equivalently reduce the atomic tunnelling rate), one
enters the hard-core boson region where each site can only have one or zero
atom due to the interaction blockade. For hard core bosons in the conventional
square lattice, with only the neighboring interaction, the supersolid state
can not be stabilized \cite{5}, and one has to rely on more unusual
interaction configuration (such as a Hamiltonian with strong
next-nearest-neighbor interaction \cite{6}) or lattice geometry (such as a
triangular lattice \cite{7}) to stabilize such a phase. A recent interesting
study based on quantum Monte Carlo simulation has shown, however, with true
long-range interaction that falls off with distance by the cubic form, the
supersolid state can be stabilized for hard-core bosons in a square lattice \cite{8}.
As the supersolid state can not be stabilized by the neighboring interaction
alone, the small energy scale associated with the long-range tail of the
dipole interaction clearly plays an important role in determining the
stability region of the supersolid state in that case.

The dipole interaction has two characteristic features: first, it is
relatively long-range; and second, the interaction form is anisotropic in
space. Ref. \cite{8} has shown that the long range tail of the dipole interaction, even under an isotropic
interaction form, can lead to stabilization of the supersolid state. Complementary to this research, in this paper we show that
by tuning the anisotropy of the dipole interaction, we can stabilize a
supersolid state even if we neglect the small long range tail of the dipole
interaction. The supersolid state here is of a different type compared with
the case studied in \cite{8} as we have a stripe instead of a checkerboard
density-wave order. The anisotropy of the dipole interaction has been analyzed
recently under a mean-field treatment of the Hamiltonian for the soft-core
bosons \cite{9}. The mean-field approximation, however, typically
overestimates the stability region of the supersolid state. For instance, it
would predict a checkerboard supersolid state for hard-core bosons in a square
lattice, which is actually unstable to quantum fluctuation \cite{5}. In this
paper, we use a well-controlled numerical method based on the recently
developed tensor network algorithm to calculate the phase diagram of hard-core
bosons in a two-dimensional square lattice. We construct quantitative phase
diagrams under different interaction parameters, and analyze the properties of
the transition from the striped solid phase to the supersolid state.
The transition has a mechanism different from the case of the
soft-core Bose Hubbard model \cite{4}.

We consider a system where bosonic dipolar molecules (or magnetic atoms) are
confined to a two-dimensional (2D) optical lattice, denoted here as $x-z$
plane. Initially the orientation of the external field is controlled such that
all the dipoles point to the $y$ axis. The strength of the dipole-dipole
interaction depends on $\theta_{ij}$ ,the angle between the orientation of
the dipole moments and their relative positions, as $(1-3 cos^{2}%
\theta_{ij})$. Thus we have isotropic repulsive interaction at the beginning
when the direction of the dipole is perpendicular to the $x-z$ plane
($\theta_{ij}=\pi/2$). Due to the strong repulsive dipole interaction, we
assume the system is in the hard-core boson region with each site occupied by
less than one molecule. Then we adiabatically tune the direction of the
external field towards the $x-z$ plane to change the anisotropy of the dipole
interaction. Depending on the angle $\theta_{ij}$, the interaction may get
attractive along one direction and repulsive along the other direction. With
the attractive interaction, one may expect that more than one bosons can
occupy the same site for the ground state. There is also concern about
stability of the system under the attractive interaction. However, as we
start with a configuration with no double occupation, the energy shift
$\Delta$ from the strong on-site interaction forbids two bosons jumping to the
same site due to the energy conservation. This is true independent of the sign
of $\Delta$, as long as its magnitude $\left\vert \Delta\right\vert $
is much larger than the atomic tunnelling rate. In addition, the possible
instability of the double occupation (or multiple occupation) states will
further suppress the probability of double occupation due to the
dissipation-induced blockade mechanism (the quantum Zeno effect) \cite{10}. As
a net result, when we tune the interaction to the attractive region, the
system remains stable in an optical lattice and still stays in the hard-core
boson region with negligible double occupation. As the dipole interaction
falls off with distance pretty quickly by the cubic law, in this paper we keep
the off-site interaction only to the order of nearest neighbors. So the focus
of investigation here is on new properties induced by the anisotropy of
the dipole interaction, not by its small long-range tale. The Hamiltonian of
this system can then be described as hard-core bosons on a 2D square lattice
with anisotropic nearest neighbor interactions%

\begin{align}
H=-\sum_{\left\langle i,j\right\rangle }  &  t(b_{i}^{+}b_{j}+h.c.)-\sum
_{i}\mu b_{i}^{\dagger}b_{i}\nonumber\\
&  +\sum_{j}(V_{x}n_{j}n_{j\pm\hat{e}_{x}}+V_{z}n_{j}n_{j\pm\hat{e}_{z}})
\end{align}
where $b_{i}^{+}$ is the boson creation operator at site $i$, $n_{i}=b_{i}%
^{+}b_{i}$, $t$ is the atomic tunnelling rate over the nearest neighbors
$\left\langle i,j\right\rangle $, $\mu$ is the chemical potential, and $V_{z}$
($V_{x}$) are respectively the nearest neighbor interaction rates in the
$z\left(  x\right)  $-direction.

The hard core bosons live in a truncated Hilbert space spanned by only two
states $\left\vert 0\right\rangle $ and $\left\vert 1\right\rangle $,
representing respectively zero or one boson on a site. The commutation
relation of the hard-core boson operators is thus modified to $\left[
b_{i},b_{i}^{+}\right]  =\left\vert 0\right\rangle $ $\left\langle
0\right\vert -\left\vert 1\right\rangle $ $\left\langle 1\right\vert
=1-2n_{i}$. With the well known mapping of the hard-core boson operators to
the Pauli operators through $b_{i}^{+}\rightarrow\sigma_{i}^{+}$ and
$2n_{i}-1\rightarrow\sigma_{i}^{z}$, the hard core boson Hamiltonian in Eq.
(1) is equivalent to the following anisotropic XXZ model
\begin{align}
H=-\sum_{\left\langle i,j\right\rangle }  &  J_{xy}(\sigma_{i}^{x}\sigma
_{j}^{x} +\sigma_{i}^{y}\sigma_{j}^{y})+V_{x}/4\sum_{j} \sigma_{j}^{z}%
\sigma_{j+\hat{e}_{x}}^{z}\nonumber\\
&  +V_{z}/4\sum_{j} \sigma_{j}^{z}\sigma_{j+\hat{e}_{z}}^{z}-h\sum_{j}%
\sigma_{j}^{z}%
\end{align}
where $J_{xy}=2t$ and $h=\mu/2-2$. The chemical potential $\mu$ acts as an
effective magnetic field along the $z$ direction in the resulting XXZ model.
With different signs of $V_{x}$ and $V_{z}$, one can have ferromagnetic
coupling in one direction and anti-ferromagnetic coupling in the other direction.

We numerically study the phase diagram of this Hamiltonian by using the iPEPS
algorithm \cite{11}. The iPEPS algorithm\ is an extension of the well-known
DMRG\ (density-matrix renormalization group) method to 2D\ quantum systems. It
is a variational approach based on the tensor network states (or called the
PEPS\ states) that appropriately capture the entanglement structure of the 2D
quantum systems. The iPEPS algorithm has been tested to work reliably for a
number of 2D many-body models \cite{12}. We have tested our numerical codes
for implementation of the iPEPS\ algorithms by comparing our calculation
results with the known results for several many-body models, including the 2D
Ising model with a transverse field and the XXZ model. In general, the results
are in quantitative agreement with the previous quantum Monte Carlo simulation
and the PEPS calculation from other groups, and the phase transition points
can be determined with a pretty good precision (the error is typically about
or less than a percent level). So we expect that with the same algorithm, we
can reliably determine the phase diagram of the anisotropic
XXZ\ model shown in Eq. (2). In our simulation, we use a bond dimension up to
$D=4$ or $5$ for the variational tensor network states and a large enough
cutoff dimension $\chi$ (typically of the order of $D^{2}$) in contraction of
the tensor network states to ensure converge of the ground state energy.

A typical phase diagram of the Hamiltonian (1) is shown in Fig 1. We take
$t=1$ as the energy unit and scan over different $\mu$ and $V_{x}$ with a
fixed $V_{z}$. To identify different phases, we calculate the superfluid order
parameter$\ \left\langle b_{i}\right\rangle $ and the boson occupation number
$\left\langle n_{i}\right\rangle $ at different sites as functions of $\mu$
and $V_{x}$. These curves show different characteristic behaviors and we can
use them to identify different phases.

Figure 1 shows the phase diagram when $V_{z}$ is fixed at $-t$ and $V_{x}$ is
scanned over the positive region. First, we note in Fig. 1 there is a large
region of the charge density wave state at exactly half filling, which has a
stripe order. This phase can be easily understood: with negative $V_{z} $ and
positive $V_{x}$, at half filling the particles arrange themselves into
stripes along the $z$ direction to maximize the neighboring interaction along
the $z$ direction and simultaneously minimize the repulsive coupling along the
$x$ direction. The most interesting feature from Fig. 1 is that the stripe
phase is always surrounded by a finite region of the supersolid state that
interconnects the superfluid phase and the charge density wave state. The
supersolid state here is characterized by coexisting of the superfluid order
and the stripe charge density wave order. The optimal condition to get the
supersolid state is to have $V_{z}$ negative and comparable with the tunnelling
rate $t$ in magnitude.

\begin{figure}[ptb]
\includegraphics[width=8cm]{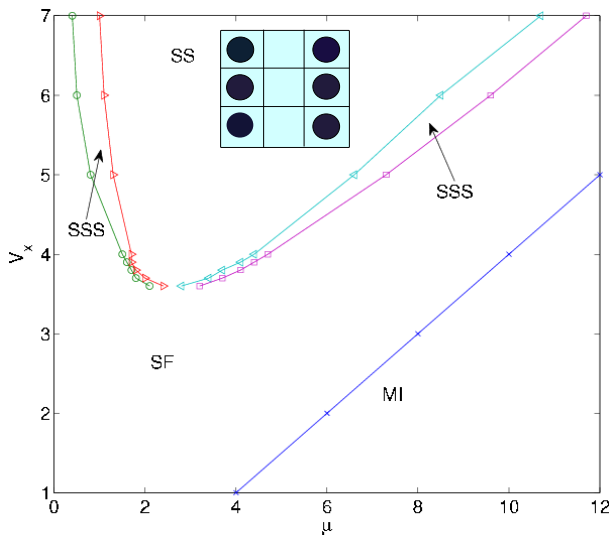}\caption[Fig. 1 ]{(Color Online) Zero-temperature
Phase diagram of the extended Bose Hubbard model with anisotropic interaction shown in Eq. (1) at $V_{z}=-t$.
As one varies the chemical potential $\mu$ and the interaction rate $V_{x}$ (both in the unit
of the hopping rate $t$), four different phases are observed,
including the superfluid (SF) phase, the striped solid (SS) phase, the striped supersolid (SSS) phase, and
the Mott insulator (MI) state. The inset shows schematically the particle filling pattern in the striped
solid phase.}%
\end{figure}

To better understand the transition to the supersolid state, in Fig. 2 we show
the evolution of the superfluid order parameter $\left\langle b_{i}%
\right\rangle $ and the particle occupation number $\left\langle
n_{i}\right\rangle $ at two alternating sites $i$ and $i+1$ along the $x$
direction as we scan the chemical potential $\mu$. Starting from a striped
solid state at half filling, we enter the supersolid phase with either
particle or hole doping. When we add particles to the half filled lattice
(with increase of the chemical potential), while the occupation of the
initially empty sites increases, the average occupation of the initially
filled sites (which form the stripes) continuously decreases. This shows that
with particle doping, some particles need to move from the initially filled
stripes to the empty sites to build up the superfluid order. As the superfluid
order increase, the charge density wave order continuously decreases and
eventually vanishes, and one enters the normal superfluid phase. The
transitions from the stripped solid phase to the supersolid and from the
supersolid to the superfluid phase are both of the second order characterized
by kinks in the first-order derivatives of the ground state energy. We have a
similar picture of the supersolid transition at the side of the hole doping
where we exchange particles with holes. This picture of the supersolid
transition is different from the mechanism of the supersolid
state for soft core bosons in a lattice with large neighboring interaction \cite{4}.
In the latter case, starting with a checkerboard lattice at half filling, the
supersolid phase only appears with particles doping (no supersolid with hole
doping), and as one add particles, these added particles continuously go to
the already filled sites to maintain the checkerboard order. As one increase
the chemical potential, the occupation of the initially filled sites
continuously increase for the checkerboard supersolid phase (instead of
decrease for the stripped supersolid phase discussed above).

\begin{figure}[ptb]
\includegraphics[width=8cm]{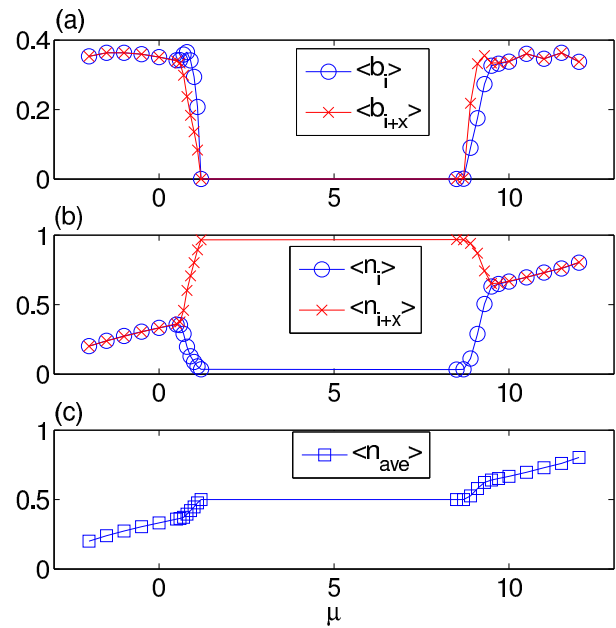}\caption[Fig. 2 ]{(Color Online) The order parameters and
the filling numbers as functions of the chemical potential. (a): The superfluid order parameters
$\langle b_{i}\rangle$ and $\langle b_{i+\hat{x}}\rangle$ at two alternating sites along
the $x$-direction (the direction perpendicular to the stripes). (b): The filling number (the particle per site)
$\langle n_{i} \rangle $ and $\langle n_{i+\hat{x}}\rangle$ at two neighboring sites along
the $x$-direction. (c): The average number per site $\langle n_{ave}\rangle = (\langle n_{i} \rangle + \langle n_{i+\hat{x}}\rangle)/2 $.
The interaction parameters in these figures are taken as $V_{x}=6t$ $V_{z}=-t$. The kinks in Figs. (a)-(c) mark two
continuous phase transitions from the superfluid to the supersolid and then to the striped solid phase.} %
\end{figure}

In the supersolid state, we can look at both the superfluid and the
density-density correlations, and these correlations are shown in Fig. 3. The
superfluid density is not homogeneous in space. In Fig. 3a, we can see that
along the $x$ direction (perpendicular to the stripe direction), the
superfluid correlation shows the zigzag pattern, but it is extended to the
long range. Along the stripe ($z$) direction, the superfluid correlation is
monotonic and approaches a constant nonzero value. This constant value,
however, is different for the particle-dominated and the hole-dominated
stripes. When the filling number is less than one half (hole doping to the
stripe solid state), the superfluid density is larger in the
particle-dominated stripes. The reverse is true for the supersolid state with
the filling number larger than one half (see Fig. 3c). The density-density
correlation is shown in Fig. (3a), which shows a long-range zigzag pattern
along the $x$ direction, but with constant correlation along the stripes ($z$
direction) as the density distribution along the stripes is homogeneous.

\begin{figure}[ptb]
\includegraphics[width=8cm]{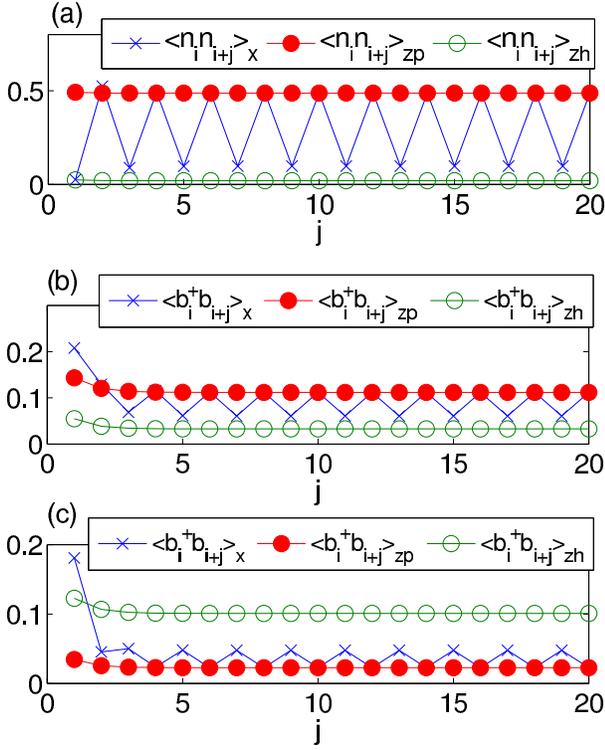}\caption[Fig. 3]{(Color Online) The density-density
and the superfluid correlations in the supersolid phase. Figs. (a) and (b) are for the
supersolid phase with hole doping to the striped solid state at half filling (with the parameters $V_{z}=-1$, $V_{x}=6$, and
$\mu=0.9$ in the unit of the hopping rate $t$). The density-density (a) and the superfluid (b) correlations are shown along the $x$ and the $z$ directions,
distinguished by the subindeces $x$ (along the x-direction), $zh$ (along the hole dominated stripe in the $z$ direction), and
$zp$ (along the particle dominated stripe in the $z$ direction). Fig. (c) represents the corresponding superfluid correlations for the
supersolid phase with particle doping to the stripe solid phase (with the parameters $V_{z}=-1$, $V_{x}=6$, and
$\mu=9.1$). One can see that with the hole doping, the superfluid correlation is stronger along the particle dominated stripe; the
reverse is true for the case of the particle doping.}
\end{figure}

In the phase diagram shown in Fig.1 , one can see that both the stripe solid
state and the supersolid phase disappear when the repulsive interaction along
the $x$ direction becomes weak. As a result, there is only the
conventional superfluid to the Mott insulator transition, and the strip phase
at half filling becomes unstable. This instability can be intuitively
understood as follows: assume we have a strip phase, and we move one particle
from the filled stripe to the empty stripe to form a particle-hole excitation.
The cost in the interaction energy is given by $2V_{x}-2V_{z}$. At the same
time, as the particle and the hole can freely move, the kinetic energy is
lowered by an amount $8t$. The net energy cost is positive when $2V_{x}%
-2V_{z}-8t>0$, and in this case the stripe phase is stable. Otherwise, the
freely-moving particle-hole pairs will be continuously generated to form a
superfluid phase and the stripe phase loses its stability. With this simple
estimate, we see that for $V_{z}=-t$, the stripe phase becomes unstable when
$V_{x}<3t$, which roughly agrees with the accurate calculation of the boundary
of the phase diagram shown in Fig. 1. This simple estimate also applies to the
cases with different $V_{z}$. For instance, in the phase diagrams shown in
Fig.4 for different values of $V_{z}$, the point where the charge wave density
(solid) phase loses its stability at half filling can still be roughly
estimated by $V_{x}<V_{z}+4t$.

\begin{figure}[ptb]
\includegraphics[width=8.5cm]{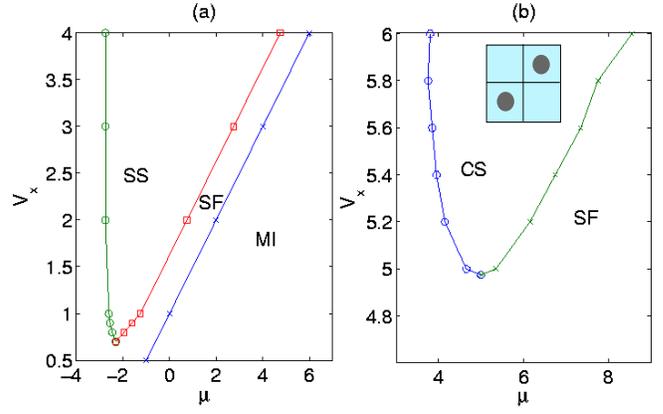}\caption[Fig. 4]{(Color Online) Phase
diagram of the Hamiltonian (1) with interaction rate $V_{z}=-3$ for Fig. (a) and
$V_{z}=0$ for Fig. (b) (in the unit of the hopping rate $t$). There is no supersolid phase in either of these two cases.
At half filling, one has a striped solid phase for Fig. (a) and a checkerboard solid
phase for Fig. (b). The insert in Fig. (b) shows schematically the particle filling pattern
in the checkerboard phase. }%
\end{figure}

As we increase or decrease the interaction $V_{z}$ away from its optimal value
around $-t$, the region of the supersolid phase gets smaller. With a stronger
attractive interaction (larger $\left\vert V_{z}\right\vert $), the stripe
solid phase gets larger in the phase diagram, however, the intermediate
supersolid phase, which requires a careful balance of the interaction energy
and the kinetic energy to enable coexistence of both the superfluid and the
charge-density wave orders, cannot be stabilized in this case. We have a
direct transition from the superfluid phase to the striped solid phase. The
evolution of the corresponding order parameter is shown in Fig. 5, which
indicates that this transition is of the first order. With a weaker $V_{z}$,
the supersolid region also gets smaller. As we turn off $V_{z}$ or scan
$V_{z}$ to the positive region, the charge-density wave state at half filling
has a checkerboard order instead of a stripe order. For the transition from
the superfluid phase to the checkerboard solid phase, we do not find any
supersolid state in the intermediate region. This finding is consistent with
the result from quantum Monte Carlo simulation of the XXZ model in a square
lattice (which corresponds to a particular case of the Hamiltonian (2) with
isotropic coupling $V_{z}=V_{x}>0$), which shows that the supersolid phase
predicted by the mean-field theory is unstable to quantum fluctuation
\cite{5}. The evolution of the order parameters as functions of the chemical potential
follow very similar curves as those shown in Fig. 5, and the transition from
the superfluid to the checkerboard phase is also of the
first-order.

\begin{figure}[ptb]
\includegraphics[width=8cm]{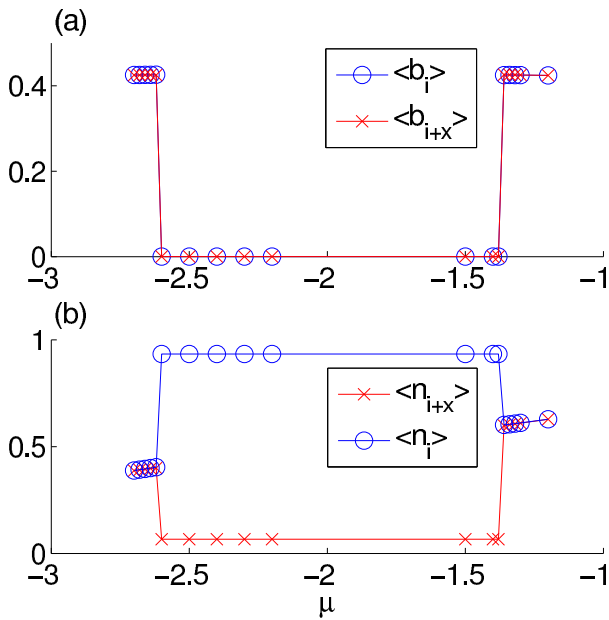}\caption[Fig. 5 ]{(Color Online) The superfulid (Fig. a) and
the charge density wave (Fig. b) order parameters are shown as functions of the chemical potential $\mu$ for the phase
diagram shown in Fig. (4a) with the interaction parameters $V_{x}=1$ $V_{z}=-3$ (in the unit of the hopping rate $t$). Notations have the same meaning as in Fig. 2.
The transition from the superfluid to the striped solid phase is clearly of the first order.
}%
\end{figure}

In summary, we have shown that anisotropy of the dipole interaction between
magnetic atoms or polar molecules can stabilize new quantum phases in an
optical lattice. By tuning the orientation of the external field, we argue
that the system can be described as a extended hard-core Bose-Hubbard model
with attractive interaction along one direction and repulsive interaction
along the other direction in a two-dimensional square lattice. Starting from
appropriate initial states, the insatiability and the atom clustering
associated with the attractive interaction can be overcome in an optical
lattice through the blockade effect induced by both the atomic interaction and
the collision loss. Using a well controlled numerical method based on the
tensor network algorithm, we calculate the phase diagram of the extended
hard-core Bose-Hubbard model with anisotropic interaction, and find a
significant region of the supersolid phase that interconnects the striped
solid phase at half filling and the conventional superfluid state. The
properties of the supersolid phase and the corresponding phase transitions are
discussed and characterized through calculation of various kinds of
correlation functions.

We thank Guin-Dar Lin for helpful discussions. This work is supported by the DARPA OLE Program under ARO Award W911NF0710576, the
AFOSR MURI program, the IARPA, and the ARO MURI program.

\end{document}